\documentclass[preprint,aps,draft]{revtex4}
\begin{document}

\title[]{Information-Theoretic Determination of Ponderomotive
Forces}

\author{Mario J. Pinheiro}
\address{Department of Physics and Centro de Fisica de Plasmas,\&
Instituto Superior Tecnico, Av. Rovisco Pais, \& 1049-001
Lisboa, Portugal} \email{mpinheiro@ist.utl.pt}



\date{\today}%
\begin{abstract}
From the equilibrium condition $\delta S=0$ applied to an isolated
thermodynamic system of electrically charged particles and the
fundamental equation of thermodynamics ($dU = T
dS-(\mathbf{f}\cdot d\mathbf{r})$) subject to a new procedure, it
is obtained the Lorentz's force together with non-inertial terms
of mechanical nature. Other well known ponderomotive forces, like
the Stern-Gerlach's force and a force term related to the
Einstein-de Haas's effect are also obtained. In addition, a new
force term appears, possibly related to a change in weight when a
system of charged particles is accelerated.
\end{abstract}
\maketitle
\section{Introduction}

There has been an effort to understand macroscopic phenomena,
characterized by thermal equilibrium and irreversible approaches
to it, based on the crucially important property: the entropy. E.
T. Jaynes~\cite{Jaynes86} was a precursor of this line of thought.
In fact, the application of information-theoretic techniques to a
number of fields has been very successful, specially in studying
colloidal systems. Interestingly, it was discovered entropy
depletion effects in a binary mixture consisting in the increase
of entropy of one component and forcing the other component to a
greater order, leading through this process to the possibility of
entropy control of the directed motion of colloidal
particles~\cite{Dinsmore96,physreve59}. Direct measurements of
entropic forces were measured for a single colloidal sphere close
to a wall in a suspension of rigid colloidal
rods~\cite{PhysRevLett03}. Also, stochastic resonance-like
phenomena in non-linear systems can be described by a maximization
of the information-theoretic distance measures between probability
distributions of the output
variable~\cite{PhysRevLett96,PhysRevLett98}.

In this paper we follow through this line of research of maximum
entropy methods, but proposing a special procedure to analyze
systems imparted with general kind of motion, both translational
and rotational. One advantage of this method is that a simple
method of calculation is devised from energy and entropy
considerations. It gives known results through the use of entropy
of a system dynamics and in specific problems could be a possible
alternative to Hamilton theory. We don't analyze the eventual deep
meaning of the present procedure, but only claim that the results
obtained mean the entropy, as it appears in the fundamental
equations of thermodynamics, could be a mathematical entity useful
for the calculation of non equilibrium processes.

Based on the maximization of entropy principle and the fundamental
equation of thermodynamics ($dU=TdS-(\mathbf{f}\cdot d
\mathbf{r})$), it was previously suggested a new procedure
~\cite{Pin02} to obtain the dynamical equation of motion of a
particle in a rotating frame.

In this paper we extend the same procedure to classical
electrodynamics, obtaining what we would like to call
information-theoretic ponderomotive forces. Besides the classical
force in a rotating frame, it is additionally obtained the
Lorentz's force acting over an electrically charged particle, the
Stern-Gerlach's force, a term related to the Einstein-de Haas's
effect, as well a new term, possibly related to a change in weight
in an accelerated frame.

\section{Extremum Principle}

Our initial theoretical framework is the one advanced by Landau
and Lifshitz~\cite{Landau}. Let us consider an isolated
macroscopic system $\mathcal{S}$ composed by N infinitesimal
macroscopic subsystems $\mathcal{S}'$ (with an internal structure
possessing a great number of degrees of freedom to ensure the
validity of the entropy concept) with $E_i$, $\mathbf{p}_i$ and
$\mathbf{J}_i$, all constituted of classical charged particles
with charge $q_i$ and mass $m_i$. The system is assumed to be
isolated and so there exist seven independent additives integrals:
the energy, 3 components of the linear momentum $\mathbf{p}$ and 3
components of the angular momentum $\mathbf{L}$.

The energetics of the charged particles is modelled on the
Maxwell-Lorentz theory of the electromagnetic field. Thus, the
internal energy $U_i$ of each subsystem moving with momentum
$\mathbf{p}_i$ in a Galilean frame of reference is given by
\begin{equation}\label{gf}
E_i = U_i + \frac{p_i^2}{2 m_i} + \frac{J_i^2}{2 I_i} - q_i V_i +
q_i (\mathbf{A}_i\cdot \mathbf{v}_i),
\end{equation}
where, besides the mechanical energy, the electrostatic and magnetic energy
($U_{m,i} \to - q_i(\mathbf{A}_i \cdot \mathbf{v}_i)$) terms are
also duly inserted.

The entropy of the system is the sum of the entropy of each
subsystems (and function of their internal energy $U$, $S=S(U)$):
\begin{equation}\label{1}
S = \sum_i ^N S_i \left( E_i - \frac{p_i^2}{2m} -\frac{J_i^2}{2
I_i} - q_i V_i + q_i(\mathbf{A}_i  \cdot \mathbf{v}_i)\right).
\end{equation}
Energy, momentum and angular momentum conservation laws must be verified for a totally isolated
system:
\begin{equation}\label{2a}
  E = \sum_{i=1}^N E_i,
\end{equation}
\begin{equation}\label{2}
  \mathbf{P} = \sum_{i=1}^N \mathbf{p}_i,
\end{equation}
and
\begin{equation}\label{3}
\mathbf{L} = \sum_{i=1}^N ([\mathbf{r_i}\times
\mathbf{p_i}] + \mathbf{J_i}).
\end{equation}
In the above equations, $\mathbf{r_i}$ is the position vector
relatively to a fixed frame of reference $\mathcal{R}$,
$\mathbf{p}_i$ is the total momentum (particle + field) and
$\mathbf{J}_i$ is the total intrinsic angular momentum of the
particle, comprising a vector sum of the electron orbital angular
momentum and the angular momentum contributed by electron spin and
nuclear spin (since the electromagnetic momentum is already
included in the preceding term through $\mathbf{p}_i$). The
exploitation of the entropy principle introduces Lagrange
multipliers from which, as we will see, ponderomotive forces are
obtained.

It is necessary to find the conditional extremum; they are set up not for the function
$S$ itself but rather for the changed function $\bar{S}$:
\begin{equation}\label{4}
\bar{S} = \sum_{i=1}^N \left\{ S_i \left[ E_i - \frac{p_i^2}{2m} -
\frac{J_i^2}{2I_i} - q_i V_i + q_i (\mathbf{A}_i \cdot
\mathbf{v}_i) \right] + (\mathbf{a} \cdot \mathbf{p_i}) +
\mathbf{b} \cdot ([\mathbf{r_i} \times \mathbf{p_i}] +
\mathbf{J_i}) \right\},
\end{equation}
where $\mathbf{a}$ and $\mathbf{b}$ are Lagrange multipliers. The conditional extremum points
are obtained for
\begin{equation}\label{g2}
\begin{array}{lr}
  \frac{\partial \bar{S}}{\partial \mathbf{r}}=0; &  \frac{\partial \bar{S}}{\partial
  \mathbf{p}}=0.
\end{array}
\end{equation}
At thermodynamic equilibrium the total entropy of the body has a
maximum constrained to supplementary conditions~\ref{2a}, ~\ref{2}
and~\ref{3}.

\section{Electrodynamic ponderomotive forces}

At this point we intend to go a step further than Landau and
Lifschitz presentation~\cite{Landau}. As we are interested in
situations in which separate parts of a system (discrete
sub-systems) are internally in equilibrium (this should apply to
continuous systems as well) small changes in extensive coordinates
$\delta X_{\mu}$ will produce small deviations $\delta
\overline{S}$ in the total entropy of the system. As is known, we
can in general define a set of intensive parameters $F_{\mu}$ in
terms of the local parameter $S(X_{\mu})$ by the equation
\begin{equation}\label{}
F_{\mu} = \frac{\partial S(X_{\mu})}{\partial X_{\mu}}
\end{equation}
with $F_{\mu}$ being a local position- and time-dependent quantity
called the affinity of the sub-system. After multiplying the
affinity by the "temperature" $T$ it is obtained the entropic
forces. This is what we seek right now and we will show that with
an appropriate procedure electrodynamic ponderomotive (body)
forces are obtained. Of course we are aware of the implications
the use of entropy in nonequilibrium  can have, but in this paper
we don't analyze the concept of probability in nonequilibrium
statistical mechanics. Quoting Guttmann~\cite{Guttmann1999}, it is
sure that this matter is still \begin{quote} ill understood...and
don't have a secure foundation.
\end{quote}

Retrospectively, the developments hereby exposed can be an
illustration of the potential use of entropy for others
applications than the conventional role played in the
determination of equilibrium states.
From the partial derivative,
$\frac{\partial \overline{S}}{\partial \mathbf{p}_i}$, it is
obtained the total (particle's + field) canonical momentum (when
putting $\mathbf{v}_{rel}=T \mathbf{a}$ and
$\mathbf{\omega}=T\mathbf{b}$, and thus giving a clear meaning to
the Lagrangian multipliers):
\begin{equation}\label{}
\mathbf{p}_i = \mathbf{p}_{rel} + m [ \mathbf{\omega} \times
\mathbf{r}_i] + q_i \mathbf{A}_i.
\end{equation}
Here, $\mathbf{p}_{rel} = m \mathbf{v}_{rel}$ is the apparent
momentum as seen in a rotating frame $\mathcal{R}'$. As it should
be, besides the mechanical terms related to translation and
rotation, it appears the electromagnetic momentum term $q_i
\mathbf{A}_i$. By the other hand, a further refinement in the
formulation of the maximum entropy method by postulating the
existence of a multiplicative factor $T/2$ associated to each
degree of freedom, allowing us from the partial derivative
$\frac{\partial \overline{S}}{\partial \mathbf{r}_i}$ to obtain
the expresion
\begin{equation}\label{}
\frac{T}{2} \frac{\partial \overline{S}}{\partial \mathbf{r}_i} =
\mathbf{\nabla}_{r_i}U_i + m_i \frac{d \mathbf{v}_i}{d t} +
\frac{1}{2} \mathbf{\nabla}_{r_i} (\mathbf{\omega} \cdot
\mathbf{J}_i)
\end{equation}
In nonequilibrium, as $T$ does not have the current meaning of
thermodynamic temperature, it is merely a parameter useful to come
through with calculations ($T$ could be a kind of kinetic
temperature).

When there is no flux of entropy ($\frac{T}{2} \frac{\partial
\overline{S}}{\partial \mathbf{r}_i}=0$) and consequently the
following relation prevails
\begin{equation}\label{class}
-\mathbf{\nabla}_{r_i} U_i^{eq} = m_i \frac{d \mathbf{v}_i}{d t} +
\frac{1}{2} \mathbf{\nabla}_{r_i} (\mathbf{\omega} \cdot
\mathbf{J}_i).
\end{equation}
Here, we denote by $U_i^{eq}$ the $ith$ sub-system internal energy
at equilibrium (since it was obtained through condition
$\frac{\partial S}{\partial r}=0$) and the first term on the
right-hand side (rhs) is the mechanical force acting on the $ith$
particle on $\mathcal{R}$ frame. The physical meaning of
Eq.~\ref{class} is that, at equilibrium, there is no production of
entropy and the force acting on over the $ith$ particle may be
found by taking the gradient of a conservative potential (which
includes the rotational energy) - in fact, an effective potential
energy, useful in dynamics to find turning points of orbits in a
central field of force. Remark that the electromagnetic field is
affected to $U_i^{eq}$ and that's why does not appear explicitly
here.

Going back to Eq.~ref{}, when considering the flux of the internal
energy, through Eq.~\ref{1} it is easy to obtain the following
full expression
\begin{displaymath}
\frac{T}{2} \frac{\partial \overline{S}}{\partial \mathbf{r}_i} =
- q_i \mathbf{\bigtriangledown}_{r_i} V_i +
q_i \mathbf{\nabla}_{r_i}(\mathbf{v}_i \cdot \mathbf{A}_i)
\end{displaymath}
\begin{equation}\label{}
+ m_i \frac{d \mathbf{v}_i}{dt} -\frac{1}{2}\mathbf{\nabla}_{r_i}
\left( \frac{J_i^2}{I_i} \right) + \frac{1}{2}
\mathbf{\nabla}_{r_i}(\mathbf{\omega} \cdot \mathbf{J}_i).
\end{equation}
It is worth to rearrange some terms in order to give more clear
physical meaning to the all expression. So, using the definition
of $\mathbf{B}$ in terms of vector potential, we have the
mathematical identity
\begin{equation}\label{}
\mathbf{\nabla}_{r_i} (\mathbf{A}_i \cdot
\mathbf{v}_i)=(\mathbf{v}_i \cdot \mathbf{\nabla}_{r_i})
\mathbf{A}_i + (\mathbf{A}_i \cdot
\mathbf{\nabla}_{r_i})\mathbf{v}_i + [\mathbf{v}_i \times
\mathbf{B}_i] + [\mathbf{A}_i \times [\mathbf{\nabla}_{r_i} \times
\mathbf{v}_i]].
\end{equation}
The cross product $(\mathbf{v}_i \cdot
\mathbf{\nabla})\mathbf{A}_i$ requires an interpretation in
physical grounds. In fact, consider a charge $q$ moving with
velocity $v$ along a given direction $x$, such that it is
displaced by $x=vt$. The Maxwell equations determine the
electromagnetic field in terms of the position of field sources
relative to the position of the charge, $A(x-v t,y,z)$. It
follows, for a planar wave, a time dependence of the form
$\frac{\partial A}{\partial t}=-v\frac{\partial A}{\partial x}$.
Taking this into account, the extremum condition $\frac{\partial
\overline{S}}{\partial \mathbf{r}_i}=0$, for example, lead us to
the relationship
\begin{equation}\label{eql1}
E_i + [\mathbf{v}_i \times \mathbf{B}_i] = \frac{1}{2q_i}
\mathbf{\nabla}_{r_i}(\frac{J_i^2}{I_i} - \mathbf{\omega} \cdot
\mathbf{J}_i) - [\mathbf{A}_i \times \mathbf{\omega}],
\end{equation}
which is a condition of charges equilibrium in a rotating frame.
The fields are defined as usually by $\mathbf{E} = -
\mathbf{\nabla} V - \frac{\partial \mathbf{A}}{\partial t}$ and
$\mathbf{B}=[\mathbf{\nabla} \times \mathbf{A}]$, with the usual
meaning. In the last expression, the electromagnetic field was
computed explicitly and so the equilibrium condition acquired a
new form. In particular, we see that a gyroscopic term plays an
important role in equilibrating the plasma.

In a closed system, the total differential of U in the variables
$r$ and $S$ is given by $dU=TdS-(\mathbf{f} \cdot \mathbf{r})$.
During the interval of time $dt$ and at a given point
($\mathbf{r}$, $S$) together with the introduction of our
postulate, we have the fundamental equation of thermodynamics in a
differential form:
\begin{equation}\label{fet}
\mathbf{\nabla}_{r_i} U_i = \frac{T}{2} \mathbf{\nabla}_{r_i} S -
\mathbf{f}_i.
\end{equation}
This is the well known fundamental equation of thermodynamics,
only differing by the introduction of a factor $\frac{T}{2}$ by
each degree of freedom. This is a necessary condition in order to
fully exploit all the theoretical predictions which are possible
to achieve through the hereby proposed procedure. In our current
perspective, the entropy provides a useful guide for the likely
outcome of a plasma. Through entropy we have access to end-points
but not to the intermediate physical processes that give birth to
a final state. Thus, applying this equation (same procedure as
presented in ~\cite{Pin02} for Newton's second law) and after some
vectorial algebra, it is easily obtained the force acting on
electrically charged particles
\begin{displaymath}
\mathbf{f}_i = q_i \mathbf{E}_i + q_i [\mathbf{v}_i \times
\mathbf{B}_i] + q_i (\mathbf{A}_i \cdot
\mathbf{\nabla}_{r_i})\mathbf{v}_i + 2 q_i [\mathbf{A}_i \times
\mathbf{\omega} ] + \frac{1}{2}\mathbf{\nabla}_{r_i}
(\mathbf{\omega} \cdot \mathbf{J}_i)
\end{displaymath}
\begin{equation}\label{}
-\frac{1}{2}\mathbf{\nabla}_{r_i}\left( \frac{J_i^2}{I_i} \right)
- \mathbf{\nabla}_{r_i} U_i.
\end{equation}
But at this stage we need to take care of the physical meaning of
the mathematical entities involved, in particular the last term -
the gradient of energy in space. It is understandable that, not
far away from equilibrium, we can always assert that
\begin{equation}\label{eq1}
-\mathbf{\nabla}_{r_i} U_i=-\mathbf{\nabla}_{r_i} U_i^{eq} +
\mathbf{f}^S.
\end{equation}
Here, we denote by $\mathbf{f}^S_i$ a force term representing a
small disturbance of a constant equilibrium state characterizing
$ith$ system. Hence, the above expression can be rewritten in the
form
\begin{displaymath}
\mathbf{f}_i = \mathbf{f}_i^M + \mathbf{f}_i^L + q_i(\mathbf{A}_i
\cdot \mathbf{\nabla}_{r_i})\mathbf{v}_i - 2 q_i [\mathbf{\omega}
\times \mathbf{A}_i]
\end{displaymath}
\begin{equation}\label{force1}
- \frac{1}{2} \mathbf{\nabla}_{r_i} \left( \frac{J_i^2}{I_i}
\right) + \mathbf{\nabla}_{r_i} (\mathbf{\omega} \cdot
\mathbf{J}_i) + \mathbf{f}_i^S.
\end{equation}
Here, $\mathbf{f}_i^M$ is the mechanical force acting over the
$ith$ particle, $\mathbf{f}_i^L=q_i \mathbf{E}_i + q_i
[\mathbf{v}_ \times \mathbf{B}_i]$ is the respective Lorentz's
force. The third term on the right-hand side of the above equation
is the rate of variation of the particle's velocity along the
potential vector acting on it. Now, notice that
$\mathbf{A}_i=\sum_j \frac{q_j \mathbf{v}_j }{c r_{ij}}$ is the
potential vector (here, in Gaussian units) at the point where the
charge $q_i$ is, as obtained by a development to first order in
$\mathbf{v}_i$ of the Li\'{e}nard-Wiechert potentials. This
solution corresponds to a retarded potential and is correct only
for $\frac{v}{c} \ll 1$, otherwise we should add a relativistic
correction. It can be easily shown that the following relation
holds
\begin{equation}\label{}
q_i (\mathbf{A}_i \cdot \mathbf{\nabla}_{r_i} )\mathbf{v}_i=-q_i
[\mathbf{\omega} \times \mathbf{A}_i].
\end{equation}
To disclose its full physical meaning after a better arrangement,
it is easily seen to correspond to a kind of gyroscopic force
\begin{equation}\label{mass}
q_i (\mathbf{A}_i \cdot \mathbf{\nabla}_{r_i} )\mathbf{v}_i =
\sum_j \frac{q_i q_j}{c r_{ij}} [\mathbf{\omega} \times
\mathbf{v}_j].
\end{equation}

Inserting Eq.~\ref{mass} into Eq.~\ref{force1} the total force
acting over the $ith$ particle can be written in the final form
\begin{displaymath}
\mathbf{f}_i = \mathbf{f}_i^M + \mathbf{f}_i^L - 2 \sum_j
m_{ij}^{em} [\mathbf{\omega} \times \mathbf{v}_j]
\end{displaymath}
\begin{equation}\label{feq}
- \frac{1}{2}\mathbf{\nabla}_{r_i} \left( \frac{J_i^2}{I_i}
\right) + \mathbf{\nabla}_{r_i} (\mathbf{\omega} \cdot
\mathbf{J}_i) + \mathbf{f}_i^{S}.
\end{equation}
We have introduced in the last equation the electromagnetic mass
of the system resulting from the interaction between charges $q_i$
and $q_j$ distant $r_{ij}$ apart
\begin{equation}\label{emass}
m_{ij}^{em} = \frac{1}{2}\frac{q_i}{c} \frac{q_j}{r_{ij}}.
\end{equation}
The third term on the rhs - a kind of gyroscopic force - is
unexpected and requires further study. It can lead to striking new
phenomena, such as levitation, because it means that the
interaction between a system of charges with different sign leads
to a mass increase, whereas equal sign charged particles when
interacting lead to a mass decrease. An effect similar to some
extent to this one was referred by Boyer~\cite{Boyer} when
investigating the change in weight associated with the
electrostatic potential energy for a system of two point charges
supported side by side against a weak downwards gravitation field.

Finally, the fourth term in the rhs is closely related to the
Einstein-de Haas effect. In fact, there is a natural relationship
between $\mathbf{J}_i$ and the Amp\`{e}rian current constituting
part of the material medium and the magnetization vector (in
particular, $\mathbf{\mu}=g\frac{e\mathbf{J}}{2mc}$), and from
this result a body force~\cite{Page40}. The fifth term in the rhs
is the Stern-Gerlach's force when applied to a magnetic moment.

The last term in the rhs, $\mathbf{f}_i^{S}$, was introduced {\it
mutatis mutandis} to represent non-equilibrium processes, such as
bremsstrahlung radiation, line radiation, turbulence, {\it inter
alia}. Altogether, in order to represent fully non equilibrium
processes, other terms should be included, such as the pressure
gradients $\mathbf{\nabla}p$ as well the dissipative terms related
to the momentum gain of the ion fluid caused by collisions with
electrons - as admitted in ~\cite{Spitzer}, they are of the form
$\mathbf{P}_{ei}=\eta e^2 n^2 (\mathbf{v}_e-\mathbf{v}_i)$, where
$\eta$ is the specific resistivity (in general, a tensor), but in
particular for an isotropic plasma it is a scalar. When searching
for a more complete description, and transposing the equations
from the discrete to the continuum level, the magnetohydrodynamics
fluid equations of motion should be retrieved.

We don't attempt to derive Maxwell equations from the actual
framework since they are essentially relativistic equations
obeying Lorentz transformations and a very careful definition of
time and space should be addressed.

\section{Conclusion}

This letter has extended the application of a new procedure to
electromagnetic fields. We believe that the development hereby
presented can show up new features related to the entropy concept
itself. It was shown that the inertial force terms appearing in
accelerated frames and the electromagnetic force acting over an
electrically charged particle are a manifestation of an entropic
force. But if the known forces are obtained with a given kind of
system energy, therefore, working backward we can conclude that a
given arrangement of the system in terms of microstates (and hence
through entropy) are generating those forces. In this sense, as
the application of the entropy concept is not well funded in
non-equilibrium situations, we have here possible example of how
far we can go. Besides the already know force terms a kind of new
electromagnetic gyroscopic force was shown to appear in a rotating
frame implying a mass variation of a system of charges.

Research of phenomena involving electromagnetic fields could take
advantage of this tool.


\end{document}